\documentclass[11pt]{article}
\usepackage{amsmath}
\usepackage{amssymb}
\usepackage{graphicx}
\usepackage{qcircuit}
\usepackage{upquote}
\usepackage{array}

\newtheorem{theorem}{Theorem}[section]

\newtheorem{proposition}[theorem]{Proposition}

\def\qed{\hfill $\Box$\medskip}

\def\cE{{\mathcal E}}
\def\ra{{\rangle}}
\def\la{{\langle}}

\def\({\left (}
\def\){\right )}

\def\diag{{\rm diag}\,}
\def\tr{{\rm tr}\,}

\def\ket#1{| #1 \rangle}

\begin{document}
\openup 1\jot

\title{Error correction schemes for fully correlated quantum 
channels protecting both quantum and classical information}

\author{Chi-Kwong Li, Seth Lyles, Yiu-Tung Poon}
\date{}
\maketitle
\begin{abstract}
We study efficient quantum error correction schemes
for the fully correlated channel on an
$n$-qubit system with error operators
that assume the form $\sigma_x^{\otimes n}$, 
$\sigma_y^{\otimes n}$,  $\sigma_z^{\otimes n}$.
Previous schemes are improved to facilitate 
implementation.
In particular, when $n$  is odd and 
equals $2k+1$, 
we describe a quantum error correction scheme 
using one arbitrary qubit $\sigma$
to protect the data state $\rho$ in a $2k$-qubit system. 
The encoding operation  $\sigma \otimes \rho 
\mapsto \Phi(\sigma \otimes \rho)$ 
only requires $3k$ CNOT gates (each with one control bit and 
one target bit). After the encoded state 
$\Phi(\sigma\otimes \rho)$ goes through
the channel, we can apply the inverse operation $\Phi^{-1}$
to produce $\tilde \sigma \otimes \rho$  so that a partial 
trace operation can recover $\rho$.
When $n$ is even and equals $2k+2$, 
we describe a hybrid quantum error 
correction scheme 
using any one of the two classical bits
$\sigma \in \{|ij\ra \la ij|: i, j \in \{0,1\}\}$
to  protect a $2k$-qubit state 
$\rho$ and 2 classical bits.
The encoding operation $\sigma \otimes \rho 
\mapsto \Phi(\sigma \otimes \rho)$
can be done by $3k+2$ CNOT gates and a 
single quibt Hadamard  gate. 
After the encoded state $\Phi(\sigma\otimes \rho)$ goes 
through the channel, we can apply the inverse operation 
$\Phi^{-1}$ to produce $\sigma \otimes \rho$  so that a perfect 
protection of the two classical bits $\sigma$ and the
$2k$-qubit state is achieved. 
If one uses an arbitrary $2$-qubit state $\sigma$, 
the same scheme will protect $2k$-qubit states.
The scheme was implemented using Matlab, Mathematica, Python, 
and the IBM's quantum computing framework 
\verb|qiskit|.
\end{abstract}

\noindent
\section{Introduction}

In quantum information processing, information is stored 
and processed with a quantum system. In the mathematical 
setting, quantum states are
represented as density matrices, i.e., complex positive 
semi-definite matrices with trace one. 
Denote by $D_N$ the set of density matrices in the set 
$M_N$ of $N\times N$ complex matrices. A qubit (quantum bit)
is a matrix,  and a quantum state of an $n$-qubit system will 
be a matrix in $D_N$ with $N = 2^n$. A quantum channel on 
an $n$-qubit system is a
trace preserving completely positive linear map
$\cE: M_N\rightarrow M_N$ that admits the operator sum 
representation \cite{C}
$$\cE(\rho) = \sum_{j=1}^r F_j \rho F_j^\dag  \qquad 
\hbox{ for all } \rho \in M_N,$$
where $\sum_{j=1}^r F_j^\dag F_j = I_N$. The matrices 
$F_1, \dots, F_r$ are
called the error operators of the quantum channel $\cE$, 
which  is the source of the
corruption of the quantum states corresponding to decoherence 
and other quantum effect
on the quantum state $\rho$. To protect the information stored 
in the quantum state $\rho$, one can use quantum 
error correction schemes to encode the
quantum state $\rho$ with some auxiliary qubits $\sigma$  
so that one can recover the quantum state $\rho$ after the 
encoded  quantum state goes through a noisy quantum channel.

Early quantum error correction scheme was designed
in analogy with classical error correction; see 
\cite{Naka,NC} and the references there in.
By adding extra ancillary qubits, one can encode 
a $k$-qubit state (data bits)
to an $n$-qubit codeword. After the codeword goes through
the noisy channel, more ancillary qubits are added 
and a syndrome measurement is performed to diagnose 
the type of error which occurred.  Based on the result of the 
error syndrome, a recovery operation is performed to recover the 
original state.
The concept of Hamming distance in classical 
coding theory was also extended to 
determine the minimum number  $n$
so that $n$-qubit codewords are needed to protect
$k$-qubit states.

In contrast to the scheme outlined in the previous paragraph,
researchers have developed new quantum error correction 
scheme in which syndrom measurement is not 
required; e.g., see 
\cite{Ket,LNPST2012} and their references.
In particular, it was shown in \cite{LNPST2012} that
for the given channel $\cE: M_N\rightarrow M_N$,
one can identify a maximum nonnegative 
integer $k \le n$, a state $\sigma \in M_{2^{n-k}}$
and  two unitary matrices  $U, V \in M_n$ so that the 
encoding and  decoding can be done as follows:

\it
For any $\rho \in D_{K}$ with $K = 2^k$, 
it can be encoded as 
$\tilde \rho = U^\dag(\sigma \otimes \rho)U$
so that $\Phi(\tilde\rho)$ has the form 
$V(\tilde \sigma \otimes \rho)V^\dag$.
Then the $\rho$ can be recovered as by taking a partial trace 
of $\tr_1(V^\dag\Phi(\tilde \rho)V) = \rho$.

\rm
The maximum number $k$ 
can be determined by the algebraic property
of the set $\{F_i^\dag F_j: 1 \le i, j \le r\}$
using the Knill-Laflamme condiion;
see \cite{KBN,LNPST2012,LNPST}.

While the new quantum error correction scheme is more efficient 
than the  previous ones, several issues needed to be addressed, 
especially, concerning   its implementation.

First, by the Knill-Laflamme condition, the given 
channel $\cE$ has a quantum error correction of dimension $K$
if and only if there is $P \in M_N$
satisfying $P^\dag = P = P^2$ with trace $K$
such that $PE_i^\dag E_j P$ is a multiple of $P$
for all $1 \le i, j \le r$; e.g., see \cite{KL,Ket,NC}.
However, checking the existence of such $P$ is highly 
non-trivial. Of course, 
the error channel may be very ``noisy'' so that there is no 
high dimensional quantum error correction code. 
Second, the encoding and decoding processes can be 
difficult to implement.  In particular, it is challenging 
to find elementary quantum gates to perform the operations
$\rho \mapsto U^\dag(\sigma \otimes \rho)U$ and 
$\Phi(\tilde \rho) \mapsto V^\dag\Phi(\tilde \rho)V$ 
followed by the partial trace operation. Although 
it is known one may use a small collection of quantum 
gates to generate all other quantum gates; 
e.g., see \cite[Chapters 4 and 6]{Naka}, 
it is tricky to
decompose specific unitary matrices $U$ and $V$ into elementary 
quantum gates which can be implemented easily.

A somewhat  ``successful example'' was considered in
\cite{LNPST}.  In the paper, 
the authors considered the fully correlated 
channel  $\cE$ on an $n$-qubit system with error operators
that assume the form $X_n =\sigma_x^{\otimes n}$, 
$Y_n = \sigma_y^{\otimes n}$, 
$Z_n = \sigma_z^{\otimes n}$, where 
$$\sigma_x = \begin{pmatrix} 0 & 1 \cr 1 & 0 \cr\end{pmatrix}, 
\quad
\sigma_y = \begin{pmatrix} 0 & -i \cr i & 0 \cr\end{pmatrix}, 
\quad
\sigma_z = \begin{pmatrix} 1 & 0 \cr 0 & -1 \cr\end{pmatrix},$$
are the Pauli matrices.
So, the quantum channel $\cE: M_N\rightarrow M_N$
with $N = 2^n$ has the form 
\begin{equation}\label{cE}
\cE(\rho) = p_0 \rho + p_1 X_n \rho X_n^\dag 
+ p_2 Y_n \rho Y_n^\dag + 
p_3 Z_n \rho Z_n^\dag \qquad \hbox{ for all } \rho \in M_N,
\end{equation}
where $p_0, p_1, p_2, p_3$ are nonnegative numbers summing up 
to one.
It was shown that if $n$ is odd, one can use a single qubit 
$\sigma \in D_2$ to protect an $(n-1)$-qubit   state;
if $n$ is even, one can use a two-qubit $\sigma$ to protect an
$(n-2)$-qubit   state. Encoding and decoding can be 
performed without measurement. Moreover, a simple recurrent
encoding and decoding scheme was presented.
As mentioned in \cite{LNPST}, although
the channel (\ref{cE}) is somewhat artificial,
the  error correction scheme may apply
in the following situation. Suppose Alice wants to
send qubits to Bob. Their qubit bases differ by unitary 
operations $X_n, Y_n$ or $Z_n$. 
Even when they do not know which basis the other party 
employs, Alice
can correctly send qubits by adding one extra qubit (when 
$n$ is odd) or two extra qubits (when $n$ is even).

In \cite{KBN}, the authors pointed out that 
the encoding scheme for the fully correlation channel
in \cite{LNPST} utilizes a pure state $\sigma$ to get
$\tilde \rho = U^\dag(\sigma \otimes \rho)U$, and it can be 
difficult and expensive to get a pure state in practice.
To address this issue, they showed that one
can use 2 arbitrary qubits to protect 1 data qubit in 
a fully correlated channel on 3-qubits.
 
In this paper, we improve the theoretical result and 
implementation schemes in several directions.

\begin{enumerate}
\item We improve the result in \cite{LNPST} and \cite{KBN}
by showing that one can use 
1 arbitrary qubit to  protect $n-1$ data qubits if $n$ is odd,
and use 2 arbitrary qubits to protect $n-2$
data qubits if $n$ is even.
\item
When $n$ is odd and equals $2k+1$, 
we describe a recursive quantum error correction scheme 
using one arbitrary qubit $\sigma$
to protect the data state $\rho$ in a $2k$-qubit system.
The encoding operation  $\sigma \otimes \rho 
\mapsto \Phi(\sigma \otimes \rho)$ 
only requires $3k$ CNOT gates (each with one control bit and 
one target bit). After the encoded state 
$\Phi(\sigma\otimes \rho)$ goes through
the channel, we can apply the inverse operation $\Phi^{-1}$
to produce $\tilde \sigma \otimes \rho$  so that a partial 
trace operation can recover $2k$-qubit state $\rho$. 
\item
When $n$ is even and equals $2k+2$, 
we describe a recursive scheme
using $2$-qubit state $\sigma$ 
to protect a $2k$-qubit state $\rho$.
The encoding operation 
$\sigma \otimes \rho 
\mapsto \Phi(\sigma \otimes \rho)$
can be done by $3k+2$ CNOT gates and a 
single quibt Hadamard  gate. 
After the encoded state $\Phi(\sigma\otimes \sigma)$ goes 
through the channel, we can apply the inverse operation 
$\Phi^{-1}$ to produce $\hat 
\sigma \otimes \rho$  so that a 
a partial  trace operation can recover $\rho$. 

\item When $n$ is even, we actually have
a hybrid quantum error correction scheme
that also protect $\sigma$ if 
$\sigma \in \{|ij\ra \la ij|: i,j \in \{0,1\}\}$.
In other words, if $\hat \rho = \sigma \otimes \rho$,
where $\sigma \in \{|ij\ra \la ij|: i,j \in \{0,1\}\}$
and $\rho \in M_{2^{n-2}}$, and 
$\Phi$ corresponds to our encoding scheme, then
$\Phi^{-1}(\cE(\Phi(\sigma\otimes \rho))) 
= \sigma \otimes \rho$.

Here we note that the study of simultaneous transmission of both quantum and classical  
information over a quantum 
channel  was initiated in \cite{DS} and followed up by other researchers,  \cite{GLZ,HW1,HW2}.

\item By the result in \cite{LNPST}, our error correction 
schemes work for any quantum channels on $n$-qubit
systems with error operators in the linear span of the set 
$\{I_{2^n}, X_n, Y_n, Z_n\}$. In fact, given a collection $\cE_1, \dots, \cE_k$
 of such channels, we can apply our
encoding scheme $\Phi$ to $\sigma \otimes \rho \in D_{2^n}$
where $\sigma \in D_2$ or $D_4$, depending on $n$ is 
odd or even, and let the encoded states
$\Phi(\sigma \otimes \rho)$  go through all these
channels a number of times. If we apply $\Phi^{-1}$ to 
the resulting state, we always get $\hat \sigma \otimes \rho$.
If $n$ is even, and $\sigma\in \{|ij\ra \la ij|:
i, j \in\{0, 1\}\}$, then $\hat \sigma = \sigma$. 
\end{enumerate}

We will state the theoretical 
results and prove them in the next section. 
Then we illustrate our schemes and depict the  circuit 
diagrams.
In Section 3, we implement and demonstrate our schemes
using Matlab, Mathematica, and IBM's  quantum computing 
framework \verb|qiskit| \cite{qiskit}.
In particular,  experiments were done
using the online quantum computers 
IBM Q 5 Yorktown and IBM Q Tenerife. 
The last section is 
devoted  to summary and discussions.

\section{The quantum error and hybrid error correction schemes}

In this section, we provide the theoretical framework 
to construct the recurrence quantum error correction schemes
mentioned in the introduction. The results are described in 
matrix theoretic setting.

For the fully correlation channel 
$\cE: M_{2^n}\rightarrow M_{2^n}$ defined in (\ref{cE}),
we construct an encoding scheme of the form 
$\rho \mapsto \Phi_n(\rho)$ such that
$$\Phi_n(\rho) = P_n \rho P_n^\dag \qquad \hbox{ for all }
\tau \in M_{2^n},$$
where $P_n \in M_{2^n}$.
One can then encode
$\rho \in D_{2^k}$ as
$P_n(\sigma \otimes \rho)P_n^\dag$, where 
$\sigma \in D_2$ or $D_4$ depending on 
$n = 2k+1$ or $2k+2$, so that 
$\Phi^{-1}(\cE(\Phi(\sigma\otimes \rho))) = \hat \sigma
\otimes \rho$.
Moreover, we have $\hat \sigma = \sigma$ if 
$n = 2k+2$ and $\sigma \in \{|q_1q_0\ra \la q_1 q_0|:
|q_0\ra, |q_1\ra \in \{|0\ra, |1\ra\}\}.$  
We will show that the matrix $P_n$
have the properties described in Section 1, namely,
\begin{itemize}
\item $P_n$ can be constructed recursively;
\item if $n$ is odd and equals
$2k+1$, then $P_n$ is a product of $3k$ CNOT
gates (with one control bit and one target bit);
\item if $n$ is even and equals 
$2k+2$, then $P_n$ is a product of
$3k+2$ CNOT gates (with one control 
bit and one target bit)
and a single qubit Hamadmard gate.
\end{itemize}

We will give detailed descriptions of the construction 
and properties of $P_n$ for $n = 2, 3$ in subsections 2.1. 
Then we describe the recurrence scheme in subsections
2.2. The general  encoding / decoding procedures
is summarized in  Section 2.3, and  
the circuit diagrams for encoding and 
decoding will be shown.

In our discussion, let 
$H = \frac{1}{\sqrt{2}} \begin{pmatrix} 1 & 1 \cr1 & -1\cr\end{pmatrix}
\in M_2$ be the Hadamard gate. 
Denote an $n$-qubit vector state by $|q_{n-1} \dots q_0\ra$. 
Let $C_{ij}$ be the CNOT gate where the $i$th qubit 
controls the target $j$th qubit.
For example, for  $(q_2,q_1, q_0) \in 
\{ (0,0,0),(0,0,1), \dots, (1,1,1)\}$, 
$$C_{02}|q_2 q_1 q_0\ra = 
\begin{cases} |q_2\oplus 1, q_1, q_0\ra 
& \hbox{ if } q_0  = 1, \cr
|q_2 q_1 q_0\ra & \hbox{ otherwise. }\cr
\end{cases}$$
In matrix form, the identity matrix in $M_8$ has 
columns 
$|000\ra, |001\ra, |010\ra, |011\ra, |100\ra, |101\ra,|110\ra, |111\ra$, the CNOT gate $C_{02} \in M_8$ has columns
$|000\ra, |101\ra, |010\ra, |111\ra, |100\ra, |001\ra,|110\ra, |011\ra$.

\subsection
{Two-qubit and three-qubit encoding/decoding operators} 

When $n = 2$, there is no quantum error correction code
for the fully correlation channel (\ref{cE}); see 
\cite{LNPST}. We will show that there is an error 
correction scheme protecting 2 classical bits,
and the scheme will provide the
basic case for the recursive quantum error correction scheme
for the fully correlation channel (\ref{cE}) on $n$-qubit systems
when $n$ is even.

Let $P_2 = C_{01} (I_2\otimes H)C_{01}^\dag\in M_4$,
where $C_{01}$ is the CNOT gate using the 
$q_0$-bit to control the $q_1$-bit for the two qubit state 
$|q_1q_0\ra$.  
In the matrix form,  
$C_{01}$ has columns $|00\ra, |11\ra, |10\ra, |01\ra$.
One readily verifies the following by direct 
calculation or any of 
the computer programs described in Section 3.

\begin{proposition} \label{2.1}
Let $P_2 = C_{01} (I_2\otimes H)C_{01}^\dag\in M_4$.
Then 
\begin{equation}\label{eq1}
(P_2^\dag X_2P_2, P_2^\dag Y_2P_2, P_2^\dag Z_2P_2) =
(D_X, D_Y, D_Z)
\end{equation}
 with 
$D_X = \diag(1,-1,1,-1), D_Y=(-1,-1,1,1), D_Z = (1, -1, -1, 1)$.
Consequently, 
$$P_2^\dag(\cE(P_2(\sigma)P_2^\dag))P_2 = \sigma$$
whenever  $\sigma = |q_1q_0\ra \la q_1q_0|$  with 
$|q_1q_0\ra \in \{|00\ra, |01\ra, |10\ra, |11\ra\}$. 
\end{proposition}

The next result concerns the encoding operator
of our error correction scheme
for the fully correlation channel (\ref{cE}) when $n= 3$.
It shows that the decomposition of the encoding operator
as the product of three CNOT gates is optimal.
Also, it provides the basic case for the recurrence scheme.
We give the short theoretical proof in the following.
The result can also be verified by the computer programs
in Section 3.

\begin{proposition}\label{2.2}
Let $P_3 =  C_{10} C_{02} C_{21} \in M_8$, where
$C_{ij}$ uses the $|q_i\ra$ to control the $|q_j\ra$
in $|q_2q_1q_0\ra$. 
Then 
 $$(P_3^\dag X_3P_3, P_3^\dag Y_3P_3, P_3^\dag Z_3P_3) = 
 (X_1 \otimes I_4, -Y_1 \otimes I_4, Z_1 \otimes I_4).$$
Consequently, if $\sigma \in D_2$ 
and $\rho \in D_{4}$, then
$$P_3^\dag(\cE(P_3(\sigma\otimes \rho)P_3^\dag))P_3 
= \tilde \sigma \otimes \rho,$$
where
$$\tilde \sigma 
= p_0 \sigma + p_1 X_1\sigma X_1^\dag + p_2 Y_1\sigma Y_1^\dag
+ p_3 Z_1 \sigma Z_1^\dag.$$ 
Moreover, $P_3$ cannot be decomposed as  a product of fewer 
than $3$ CNOT gates.
\end{proposition}

\it Proof. \rm One readily verifies the first two statements. 
For the last assertion, 
if we list the columns of $I_8$ and $P_3$ in binary form, we have
$$ I_8 = [
 \ket{000}\    \ket{001}\    \ket{010}\    \ket{011}\    
 \ket{100}\     \ket{101}\    \ket{110}\    \ket{111}],$$
 $$P_3 =  [
\ket{000}\    \ket{101}\    \ket{011}\   \ket{110}\    
\ket{111}\    \ket{010}\    \ket{100}\    \ket{001}]. 
$$
Since there are collectively 12 
mismatched positions out of the 24  positions
in the binary form of the 8 columns of the matrices 
$I_8$ to $P_3$,
and every CNOT gate will change 4 out of the 
24 positions, we see that expressing $P_3$ as the product of 
3 CNOT gates is optimal. \qed

\subsection{$n$-qubit encoding/decoding operator for 
$n \ge 4$}
 
 We first present the encoding operators for the fully
 correlation channel (\ref{cE}) when $n \ge 5$ is odd.
 
 \begin{proposition} \label{2.3} Let $n$ be an odd integer
 with $n \ge 3$. Define $P_3$ as in Proposition $\ref{2.2}$,
 and let $P_{n} 
 =(I_4 \otimes P_{n-2}) (P_3 \otimes I_{2^{n-3}})$ 
 for $n \ge 5$. Then $P_n$ is a 
 product of $3k$ CNOT gates (each has 1 
control bit and 1 target bit) with  $k = (n-1)/2$, and
$$(P_{n}^\dag X_{n}P_{n}, P_{n}^\dag Y_{n}P_{n}, 
P_{n}^\dag Z_{n}P_{n}) 
= 
 (X_1 \otimes I_{2^{n-1}}, (-1)^{k}Y_1 \otimes I_{2^{n-1}}, 
 Z_1 \otimes I_{2^{n-1}}).$$
Consequently, if $\sigma \in D_2$ and $\rho \in D_{2^{n-1}}$, 
then
$$P_n^\dag(\cE(P_n(\sigma\otimes \rho)P_n^\dag))P_n 
= \tilde \sigma \otimes \rho,$$
where
$$\tilde \sigma 
= p_0 \sigma + p_1 X_1\sigma X_1^\dag + p_2 Y_1\sigma Y_1^\dag
+ p_3 Z_1 \sigma Z_1^\dag.$$
\end{proposition}

\it Proof. \rm  By Proposition \ref{2.2}, $P_3$ is a product of 3 CNOT gates. 
 By the recursive construction, when $k$ increases by 1, we need 3 more CNOT gates.
 So, $P_n$ can be written as a product of $3k$ CNOT gates. 
 
 The other assertions can be verified readily. \qed
  
Next, we consider the case when $n$ is even. 

\begin{proposition} \label{2.4}
Suppose $n$ is an even integer with $n \ge 4$,
$P_{n-1}$ is defined as in  Proposition \ref{2.3}.
Then 
$P_{n} =(I_2 \otimes P_{n-1}) (P_2 \otimes I_{2^{n-,2}})$
is a product of $3k+2$ CNOT gates (each has 1 control bit 
and 1  target bit),
and 1 Hadamard gate (on one qubit), where $k = (n-2)/2$,
and satisfies 
$$(P_{n}^\dag X_{n}P_{n}, P_{n}^\dag Y_{n}P_{n}, 
P_{n}^\dag Z_{n}P_{n}) 
= 
 (D_X \otimes I_{2^{n-2}}, (-1)^kD_Y \otimes I_{2^{n-2}}, 
 D_Z \otimes I_{2^{n-2}})$$
$D_X = \diag(1,-1,1,-1), D_Y=(-1,-1,1,1), D_Z = (1, -1, -1, 1)$.
Consequently, if $\sigma \in D_4$ and 
$\rho \in D_{2^{n-2}}$, then
$$P_n^\dag(\cE(P_n(\sigma\otimes \rho)P_n^\dag)P_n = 
\hat\sigma\otimes \rho,$$
where 
$$\hat \sigma = p_0 \sigma + p_1 D_X \sigma D_X
+  p_2 D_Y \sigma D_Y + + p_3 D_Z \sigma D_Z.$$
In particular,
if  $\sigma 
= |q_1q_0\ra \la q_1q_0|$  with 
$|q_1q_0\ra \in \{|00\ra, |01\ra, |10\ra, |11\ra\}$,
then  $\hat \sigma = \sigma$. 
\end{proposition}

\it Proof. \rm By Proposition \ref{2.1}, $P_2$ is a product of 
2 CNOT gates and 1 Hadamard gate.  By Proposition \ref{2.3}, 
$P_{n-1}$ is a product of $3k$ CNOT gates.  So,  
$P_{n} =(I_2 \otimes P_{n-1}) (P_2 \otimes I_{2^{n-2}})$ is the product of 
$3k+2$ CNOT gates and a Hadamard gate.

The rest of the proposition can be verified readily. \qed

\subsection{The encoding/decoding schemes and circuit diagrams}

Using  
Propositions \ref{2.1} - \ref{2.4},
we can describe our encoding/decoding schemes
as follows.

\begin{theorem} \label{2.5}
Let $\cE: M_{2^n} \rightarrow M_{2^n}$
be the fully correlated quantum channel defined by
$$\cE(\rho) = p_0 \rho + p_1 X_n \rho X_n^\dag 
+ p_2 Y_n \rho Y_n^\dag + 
p_2 Z_n \rho Z_n^\dag \qquad \hbox{ for all } \rho 
\in M_{2^n}.$$
Suppose $D_X, D_Y, D_Z$,
$P_2, P_3$ and $P_n$ are defined as in Propositions
{\rm \ref{2.1}-\ref{2.4}.}

\begin{itemize}
\item[{\rm (a)}] Suppose $n$ is an odd integer
and equals $2k+1 \ge 3$. Then 
$P_n$ is a product of $3k$ CNOT gates
(each has 1 control and 1 target bit). One can 
encode an $(n-1)$-qubit data state $\rho$ using an 
arbitrary qubit $\sigma$ by the encoding 
operator $P_n$ so that
$$\rho \mapsto P_n(\sigma \otimes \rho)P_n^\dag.$$
After the encoded state goes through the fully correlated
channel $\cE$, one can apply the operation 
$\tau \mapsto P_n^\dag \tau P_n$. Then the encoded state 
$P_n(\sigma\otimes \rho)P_n^\dag$ becomes    
$$P_n^\dag\cE(P_n(\sigma\otimes \rho)P_n^\dag)P_n =
    \tilde \sigma \otimes \rho,$$
where
$$\tilde \sigma 
= p_0 \sigma + p_1 X_1\sigma X_1^\dag + p_2 Y_1\sigma Y_1^\dag
+ p_3 Z_1 \sigma Z_1^\dag.$$ 
  Applying 
a partial trace $\tr_1(\tilde \sigma \otimes \rho)$, one can
  recover the data state $\rho$.
\item[{\rm (b)}] Suppose $n$ is an even integer
and equals  $2k+2 \ge 4$.
Then $P_n$ is a product of $3k+2$ CNOT gates (each has 
1 control and 1 target bit) and a Hadamard gate (on 1 qubit).
One can  encode an $(n-2)$-qubit data state $\rho$ 
by an arbitrary $\sigma \in D_4$
using the encoding operator $P_n$ so that
$$\rho \mapsto P_n(\sigma \otimes \rho)P_n^\dag.$$
After the encoded state goes through the fully correlated 
channel $\cE$, one can apply the operation 
$\tau \mapsto P_n^\dag \tau P_n$. 
Then the encoded state becomes
$$P_n^\dag\cE(P_n(\sigma\otimes \rho)P_n^\dag)P_n =
    \hat \sigma \otimes \rho,$$
where 
$$\hat \sigma = p_0 \sigma + p_1 D_X\sigma D_X 
+ p_2 D_Y \sigma D_Y + p_3 D_Z \sigma D_Z \in D_4.$$
Applying 
a partial trace $\tr_1(\tilde \sigma \otimes \rho)$, one can
  recover the data state $\rho$.
Furthermore, if  $\sigma \in \{|ij\ra \la ij|:
 i,j \in\{0,1\} \}$, then 
$\hat \sigma = \sigma$ and the whole product state $ \sigma \otimes \rho$ is protected through the fully correlated 
channel $\cE$.
\end{itemize}
\end{theorem}

In the following,
we depict the circuit diagram of the quantum error correction
scheme described in Theorem \ref{2.5}.

\smallskip\noindent
For $n = 2$, if $|q_1q_0\ra \in \{|00\ra, |01\ra, |10\ra, |11\ra\}$, then 
circuit diagram will be:
\begin{equation*}
    \Qcircuit @C=1em @R=1.0em @!R {
        \lstick{\ket{q_0}} & \ctrl{1} & \gate{H} & \ctrl{1} &  \multigate{1}{~~~~\cE~~~~} & 
        \ctrl{1} & \gate{H} & \ctrl{1} & \qw & \rstick{\ket{q_0}}\\
        \lstick{\ket{q_1}} & \targ    &      \qw & \targ    & \ghost{~~~~\cE~~~~} & \targ & 
        \qw      & \targ    & \qw & \rstick{\ket{q_1}}
        }
    \end{equation*}
        
\medskip\noindent
For $n = 3$, 
    \begin{equation*}
    \Qcircuit @C=1em @R=1.0em @!R {
        \lstick{\ket{q_0}} & \qw & \ctrl{2} & \targ  & \multigate{2}{~~~~\cE~~~~} & \targ & \ctrl{2} & \qw & \qw & \rstick{\ket{q_0}}\\
        \lstick{\ket{q_1}} & \targ & \qw & \ctrl{-1} &\ghost{~~~~\cE~~~~}& \ctrl{-1} & \qw & \targ & \qw &\rstick{\ket{q_1}}\\
        \lstick{\ket{q_2}} & \ctrl{-1} & \targ & \qw & \ghost{~~~~\cE~~~~} & \qw & \targ & \ctrl{-1} & \qw &\rstick{\ket{\hat{q_2}}}}
    \end{equation*}
   
\medskip\noindent
For odd $n$, the circuit diagram will be:
\begin{equation*}
\Qcircuit @C=1em @R=1.0em @!R {
    \lstick{\ket{q_{0}}}   & \qw & \qw      & \qw      &\multigate{2}{P_{n-2}}       &\multigate{4}{{ ~~~~\cE~~~~ }} & \multigate{2}{P^\dag_{n-2}}      & \qw      &\qw       &\qw &\qw &\rstick{\ket{q_{0}}}\\
    \lstick{\vdots}       &\qw          &\qw       & \qw      &\ghost{P_{n-2}}       &\ghost{~~~~\cE~~~~}        & \ghost{P_{n-2}}      & \qw      &\qw     &\qw & \qw &\rstick{\vdots}\\
    \lstick{\ket{q_{n-3}}} &\qw& \ctrl{2} & \targ  &\ghost{P_{n-2}} & \ghost{~~~~\cE~~~~} & \ghost{P_{n-2}} & \targ & \ctrl{2} & \qw &\qw &\rstick{\ket{q_{n-3}}}\\
    \lstick{\ket{q_{n-2}}} & \targ & \qw & \ctrl{-1}&\qw &\ghost{~~~~\cE~~~~} & \qw & \ctrl{-1} & \qw & \targ & \qw &\rstick{\ket{q_{n-2}}} \\
    \lstick{\ket{q_{n-1}}} & \ctrl{-1} & \targ & \qw & \qw & \ghost{~~~~\cE~~~~} &\qw & \qw & \targ & \ctrl{-1} & \qw &\rstick{\ket{\hat q_{n-1}}}
    }
\end{equation*}

\medskip\noindent
For even $n$, 
if $|q_{n-1}q_{n-2}\ra \in \{ |00\ra, |10\ra, |01 \ra, |11\ra\}$,
then the circuit diagram will be:
\begin{equation*}
\Qcircuit @C=1em @R=1.0em @!R {
    \lstick{\ket{q_{0}}}    & \qw      & \qw      &\qw       & \multigate{2}{P_{n-1}} &\multigate{3}{~~~~\cE~~~~} & \multigate{2}{P^\dag_{n-1}}&\qw      & \qw      &\qw       & \qw & \rstick{\ket{q_0}}\\
    \lstick{\vdots}               &\qw       & \qw      &\qw        &\ghost{P_{n-1}}  &\ghost{~~~~\cE~~~~}  &\ghost{P^\dag_{n-1}}   & \qw      & \qw      &\qw     &\qw &  \rstick{\vdots}\\
    \lstick{\ket{q_{n-2}}}   & \ctrl{1} & \gate{H} & \ctrl{1}  &\ghost{P_{n-1}} &\ghost{~~~~\cE~~~~}  &\ghost{P^\dag_{n-1}}   & \ctrl{1} & \gate{H} & \ctrl{1}    &\qw &\rstick{\ket{q_{n-2}}}\\
    \lstick{\ket{q_{n-1}}} & \targ    &      \qw & \targ & \qw   &\ghost{~~~~\cE~~~~}  & \qw   & \targ    & \qw      & \targ    &\qw   &\rstick{\ket{q_{n-1}}}
    }
\end{equation*}

\section{Implementation}

\subsection{Matlab}

A Matlab program is written to generate the matrices $X_n, Y_n, Z_n, P_n$,
etc.,  and implement the error correction scheme
described in Section 2.

For an integer $n > 1$, the following commands will generate the encoding 
matrix $P_n$:
\begin{verbatim}
if mod(n,2) == 1   
       P = eye(8); P3 = P(:,[1,6,4,7,8,3,5,2]); Pn = P3;
          k = (n-1)/2;
       for j = 2:k
            Pn = kron(eye(4),Pn)*kron(P3,eye(2^(2*j-2)));
   end 
else 
     H = [1 1; 1 -1]/sqrt(2); C01 = [1 0 0 0; 0 0 0 1; 0 0 1 0; 0 1 0 0]; 
                    P2 = C01*kron(eye(2),H)*C01;   
   if n == 2
              Pn = P2;
   else
       P = eye(8); P3 = P(:,[1,6,4,7,8,3,5,2]); Pn = P3; k = (n-2)/2;
     for j = 2:k
           Pn = kron(eye(4),Pn)*kron(P3,eye(2^(2*j-2)));
     end  
           Pn = kron(eye(2),Pn)*kron(P2,eye(2^n/4));
   end
end
\end{verbatim}
Then one can test the encoding and decoding schemes. First, set up the error operators
$X_n, Y_n, Z_n$. 
\begin{verbatim}
%%%   Set up the error operators for the channel
     X = [0 1; 1 0];  Y = [0 -i; i 0];   Z = [1 0; 0 -1];  
        Xn=X; Yn = Y; Zn = Z;
   for j = 2:n
        Xn = kron(X,Xn); Yn = kron(Y,Yn); Zn = kron(Z,Zn);
   end
\end{verbatim}
Suppose $n = 2k+1$ is odd. The following commands check 
\begin{equation} \label{odd-a}
(P_n^\dag X_nP_n, P_n^\dag Y_nP_n, P_n^\dag Z_nP_n) = 
(X\otimes I, (-1)^k Y \otimes I, Z\otimes I).
\end{equation}
The output 0,0,0 will confirm the equality.
\begin{verbatim}
%%%  Check (Pn'XnPn, Pn'YnPn, Pn'ZnPn)
    II = eye(2^(n-1));  norm(Pn'*Xn*Pn -kron(X,II)), 
    norm(Pn'*Yn*Pn - (-1)^k*kron(Y,II)), norm(Pn'*Zn*Pn -kron(Z,II))
\end{verbatim}
Next, we verify the error correction scheme for random input $\sigma \otimes \rho$
with $\sigma \in D_2$ and $\rho \in D_4$. The output 0,0,0 will confirm the scheme works.
\begin{verbatim}
%%% Generate random S in D_2, S in D_{2k}
   S = rand(2,2) + i*rand(2,2) - rand(1,1)*(1+i)*eye(2); S = S*S'; S = S/trace(S); 
   K = 2^(n-1);    
   R = rand(K,K) + i*rand(K,K) - rand(1,1)*(1+i)*eye(K); R = R*R'; R = R/trace(R);
%  Encode kron(S,R) and compared with the decoded state for each error operator.
   A = kron(S,R); AA = Pn*A*Pn'; 
   norm( Pn'*Xn*AA*Xn*Pn - kron(X*S*X',R)), norm(Pn'*Yn*AA*Yn*Pn - kron(Y*S*Y',R))
   norm( Pn'*Zn*AA*Zn*Pn - kron(Z*S*Z',R))
\end{verbatim}
Suppose $n$ is even. The following commands check
\begin{equation}\label{even}
(P_n^\dag X_n P_n,P_n^\dag Y_n P_n,P_n^\dag Z_n P_n) = 
(D_Z \otimes I_{2^{n-2}},D_Y \otimes I_{2^{n-2}}, D_Z \otimes I_{2^{n-2}}).\end{equation}
The output 0,0,0 will confirm the equality.
\begin{verbatim}
    Dx = diag([1 -1 1 -1]); Dy = diag([-1 -1 1 1]); Dz = diag([1 -1 -1 1]);
    II = eye(2^n/4);  norm(Pn'*Xn*Pn - kron(Dx,II)); 
     norm(Pn'*Yn*Pn - (-1)^k*kron(Dy,II)), norm(Pn'*Zn*Pn - kron(Dz,II))
\end{verbatim}
Then we verify our error correction scheme that for any $\sigma \in \{|00\ra\la 00|,
|01\ra\la 01|,|10\ra\la 10|,|11\ra\la 11|\}$ and $\rho \in D_{2^{n-2}}$,
the encoding and decoding yield $\sigma \otimes \rho$. Again, the output 0,0,0
will confirm the scheme works.
\begin{verbatim}
%%%  Set up the classical bits in D_4, and arbitrary qubits in D_{2k}
   K = 2^(n-2); 
   R = rand(K,K) + i*rand(K,K) - rand(1,1)*(1+i)*eye(K); R = R*R'; R = R/trace(R);
   b0 = [1 0; 0 0]; b1 = [0 0; 0 1]; 
   b00 = kron(b0,b0); b01 = kron(b0,b1); b10 = kron(b1,b0); b11 = kron(b1,b1);
%
   S = b00; A = kron(S,R); AA = Pn*A*Pn';  norm(Pn'*Xn*AA*Xn*Pn - kron(S,R)),
       norm(Pn'*Yn*AA*Yn*Pn - kron(S,R)), norm(Pn'*Zn*AA*Zn*Pn - kron(S,R))
%
   S = b01; A = kron(S,R); AA = Pn*A*Pn'; norm(Pn'*Xn*AA*Xn*Pn - kron(S,R)), 
       norm(Pn'*Yn*AA*Yn*Pn - kron(S,R)), norm(Pn'*Zn*AA*Zn*Pn - kron(S,R))
%
   S = b10; A = kron(S,R); AA = Pn*A*Pn'; norm(Pn'*Xn*AA*Xn*Pn - kron(S,R)), 
       norm(Pn'*Yn*AA*Yn*Pn - kron(S,R)), norm(Pn'*Zn*AA*Zn*Pn - kron(S,R))
%
   S = b11; A = kron(S,R); AA = Pn*A*Pn'; norm(Pn'*Xn*AA*Xn*Pn - kron(S,R)), 
       norm(Pn'*Yn*AA*Yn*Pn - kron(S,R)), norm(Pn'*Zn*AA*Zn*Pn - kron(S,R))
\end{verbatim}

\subsection{Mathematica}

We write a Mathematica program to generate the matrices 
$X_n, Y_n, Z_n, P_n$, etc.,  and demonstrate our quantum error 
correction scheme described in Section 2.   

We briefly describe our program in the following.
We begin by setting up the CNOT gates, the Hadamard matrix, the Pauli matrices, $D_X,D_Y,D_Z$ and $P_2$ by the
following commands:

\begin{verbatim}
CNOT[n0_,h0_,k0_]:=Module[{n=n0,h=h0,k=k0},U=IdentityMatrix[2^n];
Do[cindex=IntegerDigits[i-1,2,n];
If[cindex[[n-h]]==1,cindex[[n-k]]=Mod[cindex[[n-k]]+1,2]];
s=Sum[cindex[[r]]*2^(n-r),{r,1,n}]+1;
U[[i]]=Table[KroneckerDelta[s,j],{j,1,2^n}],{i,1,2^n}];U]

H={{1,1},{1,-1}}/Sqrt[2];
x={{0,1},{1,0}};
y={{0,-I},{I,0}};
z={{1,0},{0,-1}};
DX = DiagonalMatrix[{1, -1, 1, -1}];
DY = DiagonalMatrix[{-1, -1, 1, 1}]; 
DZ = DiagonalMatrix[{1, -1, -1, 1}];
P2=CNOT[2,0,1].KroneckerProduct[IdentityMatrix[2],H].Transpose[CNOT[2,0,1]];
\end{verbatim}

\noindent Then we define $X_n$, $Y_n$ and $Z_n$ recursively:
\begin{verbatim}
X[n_]:=X[n]=KroneckerProduct[X[n-1],x]; X[1]=x;
Y[n_]:=Y[n]=KroneckerProduct[Y[n-1],y]; Y[1]=y;
Z[n_]:=Z[n]=KroneckerProduct[Z[n-1],z]; Z[1]=z;
\end{verbatim}

\noindent Then we define $P_n$ recursively by setting $P_{2k+1}=Q[k]$    
and $P_{2k}=R[k]$.
\begin{verbatim}
Q[k_]:=Q[k]=KroneckerProduct[IdentityMatrix[4],Q[k-1]]. 
KroneckerProduct[Q[1],IdentityMatrix[2^(2k-2)]];
Q[1]=CNOT[3,1,0].CNOT[3,0,2].CNOT[3,2,1];

R[k_]:=KroneckerProduct[IdentityMatrix[2],Q[k-1]]. 
KroneckerProduct[P2,IdentityMatrix[2^(2k-2)]];
\end{verbatim}
Then the validity of the formula
$$(P_{2k+1}^t X_{2k+1}P_{2k+1}, P_{2k+1}^t Y_{2k+1}P_{2k+1}, 
P_{2k+1}^t Z_{2k+1}P_{2k+1}) 
= 
 (X_1 \otimes I_{2^{2k}}, (-1)^{k}Y_1 \otimes I_{2^{2k}}, 
 Z_1 \otimes I_{2^{2k}}).$$
can be checked by calculating $\|P_n^tX_nP_n -X_1 \otimes I_{2^{2k}}\|$, 
$\|P_n^tY_nP_n -(-1)^{k}Y_1 \otimes I_{2^{2k}}\|$, $\|P_n^tZ_nP_n -Z_1 \otimes I_{2^{2k}}\|$ 
with the corresponding functions:
\begin{verbatim}
Norm[Transpose[Q[3]].X[7].Q[3] - KroneckerProduct[x,IdentityMatrix[2^6]]]
Norm[Transpose[Q[3]].Y[7].Q[3] - (-1)^3*KroneckerProduct[y,IdentityMatrix[2^6]]]
Norm[Transpose[Q[3]].Z[7].Q[3] - KroneckerProduct[z,IdentityMatrix[2^6]]]
\end{verbatim}

\noindent Similarly, we can check the formula
$$(P_{2k}^t X_{2k}P_{2k}, P_{2k}^t Y_{2k}P_{2k}, 
P_{2k}^t Z_{2k}P_{2k}) 
= 
 (D_X \otimes I_{2^{2k-2}}, (-1)^{k-1}D_Y \otimes I_{2^{2k-2}}, 
 D_Z \otimes I_{2^{2k-2}} $$
by calculating (for $k=3$) respectively as follows:
\begin{verbatim}
Norm[Transpose[R[3]].X[6].R[3] - KroneckerProduct[DX,IdentityMatrix[2^4]]]
Norm[Transpose[R[3]].Y[6].R[3] - (-1)^2*KroneckerProduct[DY,IdentityMatrix[2^4]]]
Norm[Transpose[R[3]].Z[6].R[3] - KroneckerProduct[DZ,IdentityMatrix[2^4]]]
\end{verbatim}

\noindent We can also check that for all $n\ge 3,\ \|P_n^\dag \cE( P_n (\sigma \otimes \rho) P_n^\dag) P_n  
-\tilde \sigma \otimes \rho\|=0$.

\subsection{Python}
Similar to the other methods, we can verify the results in Python. 
This is convenient because IBM has provided the \verb|qiskit| framework, and the encoding matrices can be extracted from a built-in circuit using the unitary backend \cite{qiskit}.

In contrast to the matrix operations of sections 3.1 and 3.2, the recursive scheme operates on the quantum circuit itself. For convenience, the imports have been omitted, but can be found in the full code.

To initialize a quantum circuit, we first create qubits using \verb|qr = QuantumRegister(n)| and classical bits for measurement with \verb|cr = ClassicalRegister(n)|. The quantum circuit is constructed by the \verb|qc = QuantumCircuit(qr, qc)|. To verify functionality, we can apply arbitrary unitary operations to initialize the state to a given vector. Then, the encoding scheme is defined recursively:

\begin{verbatim}
    def build_circ(qc, q, E):
        n = q[-1][1] + 1 # number of qubits

        def err(e, base=False):
            d = {'X':qc.x, 'Y':qc.y, 'Z':qc.z, 'I':lambda x: None} # apply errors
            d[e](q[n-1])
            if base: # if we're at q3
                d[e](q[0])
                d[e](q[1])
        
        if n % 2 == 0: # qn is even
            qc.cx(q[n-2], q[n-1]) # encode with P_2
            qc.h(q[n-2])
            qc.cx(q[n-2], q[n-1])
            build_circ(qc, q[:-1], E) #recurse
            err(E)
            qc.cx(q[n-2], q[n-1]) # decode with P_2^T
            qc.h(q[n-2])
            qc.cx(q[n-2], q[n-1])
        else: # it's odd
            qc.cx(q[n-1], q[n-2]) # encode with P_3
            qc.cx(q[n-3], q[n-1])
            qc.cx(q[n-2], q[n-3])
            if n == 3: # base case
                err(E, True)
            else:
                build_circ(qc, q[:-1], E) #recurse
                err(E)
            qc.cx(q[n-2], q[n-3]) # decode with P_3^T
            qc.cx(q[n-3], q[n-1])
            qc.cx(q[n-1], q[n-2])
        return()
\end{verbatim}

We can use the IBM's \verb|qasm| quantum computer simulator to validate our scheme. The results are identical to the above two sections, so we omit them here. The \verb|qasm| computer is deterministic, so an input $|q_2\ra = U|0\ra$ and  $|q_1q_0\ra = V|00\ra$ for unitary $U \in M_2, V \in M_4$ returns a perfect decoding. Similar statistics may be obtained for $n = 4,5,6$, etc.

\subsection{IBM Quantum Computer}

We use IBM's online quantum computers to  implement our scheme. We compare two different machines: Tenerife (\verb|ibmqx4|, pink in graphs) and Yorktown (\verb|ibmqx2|, blue) to show the discrepancy of prediction quality between the two. Yorktown has more than 9 times the gate error of Tenerife, and 2.4 times the readout error according to the documentation 
at \cite{ERR}.  Table 1 mostly supports these facts; however, there are some interesting results when the  input is 10 (on $Z$ and $I$ error),  where Yorktown outperforms.

We have three experimental results: $\sigma = 0$ in Table 1, 
$\sigma = 1$ in Table 2, and $\sigma$ is a randomly generated 
state in Table 3.

For $\sigma = 0$, it is interesting to observe that for inputs 
$00, 01$ Tenerife is significantly better at preserving states than Yorktown. For both of the computers, most of the error seems to take the form of $q_1$ flipping. In the case of 01 on Yorktown, the output is evenly split between the correct state and the bit flipped  state 11.

Also of interest is the fact that, empirically, the accuracy is dependent on 
$\sigma$. In the case of random $\sigma$, Yorktown is essentially ineffective 
at maintaining the state with the given encoding.

We can test the encoding scheme on $\ket{0000}$ and $\ket{00000}$, as shown 
in Figure 1. In the case of 4 qubits, Tenerife finds the correct output, but 
neither computer could produce  useful results in the 5 qubit case.

\section{Concluding remarks and future work}

We obtain an efficient error correction  
scheme for fully correlated quantum channels
protecting both quantum and classical information.
The scheme was implemented using Matlab, Mathematica, Pythons, and
the IBM's quantum computing framework \verb|qiskit|.
The codes are available at:
\begin{verbatim}
                    https://github.com/slyles1001/QECC
\end{verbatim}
Computational results were described in Section 3.4
and summarized in Tables 1-3 and Figure 1.

\medskip
Several remarks are in order concerning the implementation.

In Matlab, we generate the encoding operation $P_n$ for $n \ge 2$, and 
check  the properties described in Propositions \ref{2.1} -- \ref{2.4}.
Note that when the dimension is high, the matrices $P_n, X_n$, etc.\ 
are too big to display,
and there are too many entries to check (though most of them are zeros).
The command \verb|norm(Pn'*Zn*AA*Zn*Pn - kron(S,R))| computes the norm
of the matrix $P_n^\dag Z_n P_n(\sigma \otimes \rho)P_n^\dag Z_n P_n 
- \sigma \otimes \rho$
to confirm that it gives zero (up to machine error). In fact, 
in the odd case, the Hadamard gate 
is not used to do encoding and decoding, the norm 
values of the relevant matrices are exactly 0; in the even case, 
the  Hadamard gate  is used (once in encoding and 
once in decoding), the norm  value of matrices  yields 
a number at the order of the machine error.

In Mathematica, the norm of the matrix 
$P_n^\dag \cE( P_n (\sigma \otimes \rho) P_n^\dag) P_n  
-\tilde \sigma \otimes \rho$ is always exactly 0 even if the 
Hadamard gate is used in the even case.
Here 
$\tilde \sigma  = p_0 \sigma + p_1 X\sigma X^\dag 
+ p_2 Y \sigma Y^\dag 
+ p_3 Z \sigma Z^\dag \in D_2$ if $n$ is odd and 
$\sigma = \tilde \sigma \in D_4$ is one of the 4 classical 
binary bits if  $n$ is even. This is due to Mathematica 
being an algebraic solver versus Matlab and Python.

In the IBM quantum computer setting, it is interesting to note 
that for $U \in \{I_2, X, Y, Z\}$, when we apply the 
encoding scheme to 3-qubit
$P_3 |q_2q_1q_0\ra$, then the error operator 
$U^{\otimes 3}P_3 |q_2 q_1 q_0\ra$, and then the operator
$P_n^\dag U^{\otimes 3}P_3 |q_2 q_1 q_0\ra$, the second
qubit always attracts more error compared with the expected 
output $| Uq_2 q_1q_0\ra$,
even for the case when $U = I_2$. (See Tables 1 and 2.)
We are curious to know why such a noise
pattern is observed when implemented, but cannot speculate at 
this time.

\medskip
For future research, we plan to extend the techniques to more general 
quantum channels 
such as the fully correlated quantum channels on $n$-qubits
with general noise of the form $U^{\otimes n}$, 
where $U \in M_2$
is unitary, or a non-classical bit. We also plan to 
further investigate the cause of the systemic errors in flipping 
the most significant bit of our systems.

\medskip
Miguel A. Martin-Delgado pointed out that there were study 
and experiments
on quantum error  correction code in a  NISQ quantum computer; 
see \cite{MD1,MD2}. We would like to implement our scheme
to quantum channels on more qubits  
using other quantum computers which are not as noisy as the
IBM quantum computers.

\section*{Acknowledgments}

The authors would like to thank the referees for some 
helpful comments.
Li is an affiliate member of the Institute for Quantum 
Computing,
University of Waterloo. 
His research was supported by USA NSF grant DMS 1331021,
Simons Foundation Grant 351047.

\noindent
(C.K. Li) Department of Mathematics, College of William \& Mary,
Williamsburg, VA 23185, USA. Email: ckli@math.wm.edu

\noindent
(S. Lyles) Department of Mathematics, College of William \& Mary,
Williamsburg, VA 23185, USA. Email:
smlyles@email.wm.edu

\noindent
(Y.T. Poon) Department of Mathematics, Iowa State University,
Ames, IA 50011, U.S.A.\newline 
Center for Quantum Computing, Peng Cheng Laboratory, Shenzhen, 518055, China\newline Email: ytpoon@iastate.edu

\newpage
\
\vskip -.3in
\centerline{\includegraphics[height=4.3in,width=6.8in]{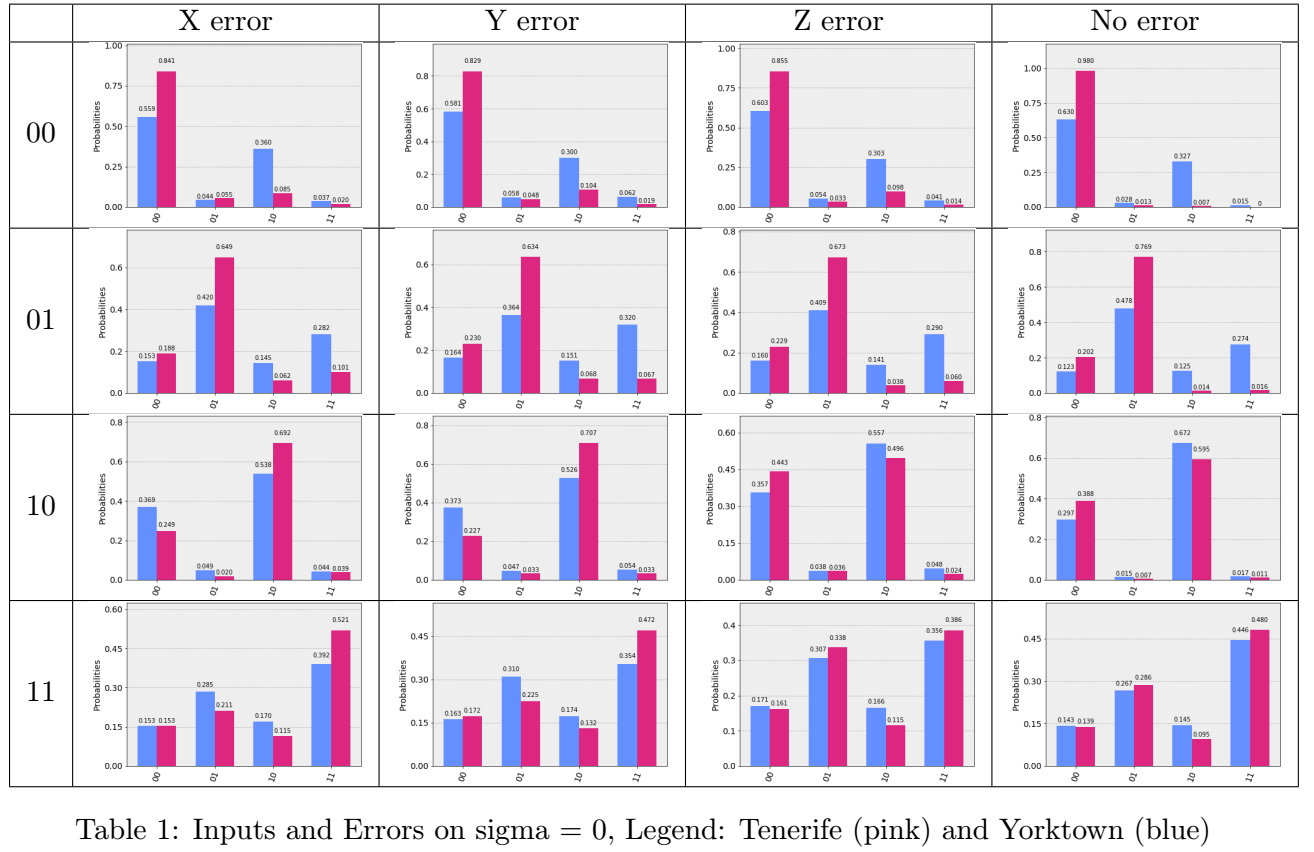}}

\bigskip
\centerline{\includegraphics[height=4.3in,width=6.8in]{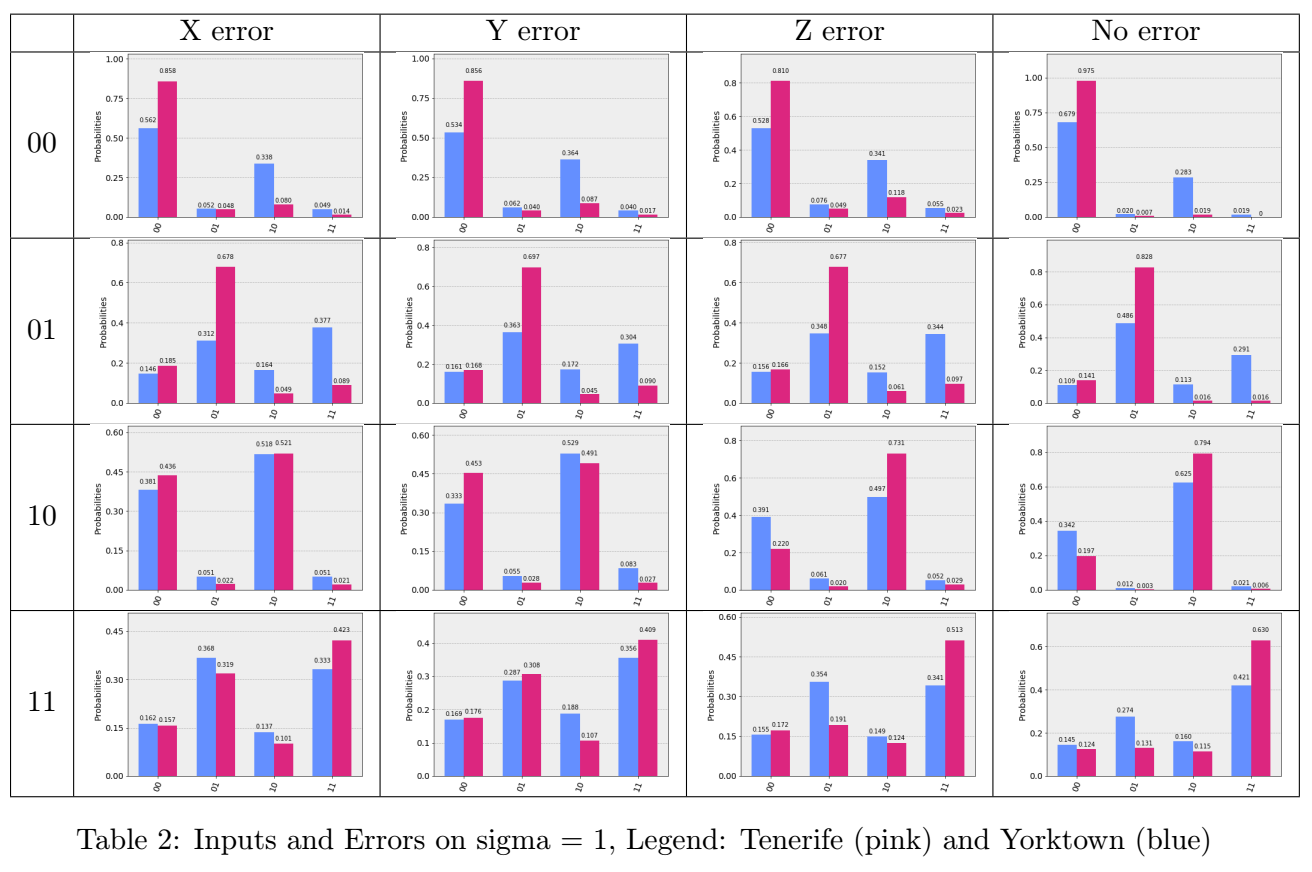}}

\newpage
\centerline{\includegraphics[height=4.3in,width=6.8in]{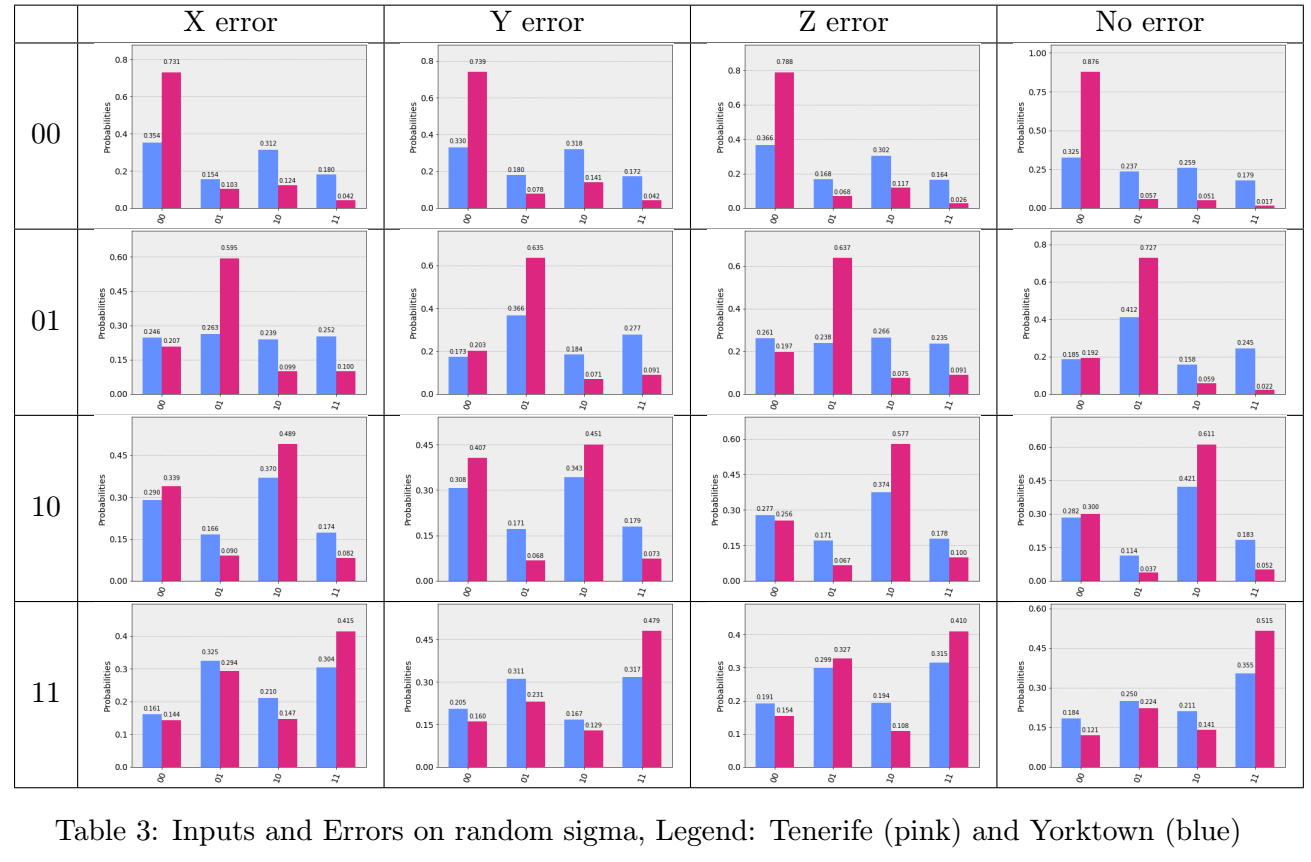}}

\bigskip
\centerline{\includegraphics[height=2.3in,width=6.8in]{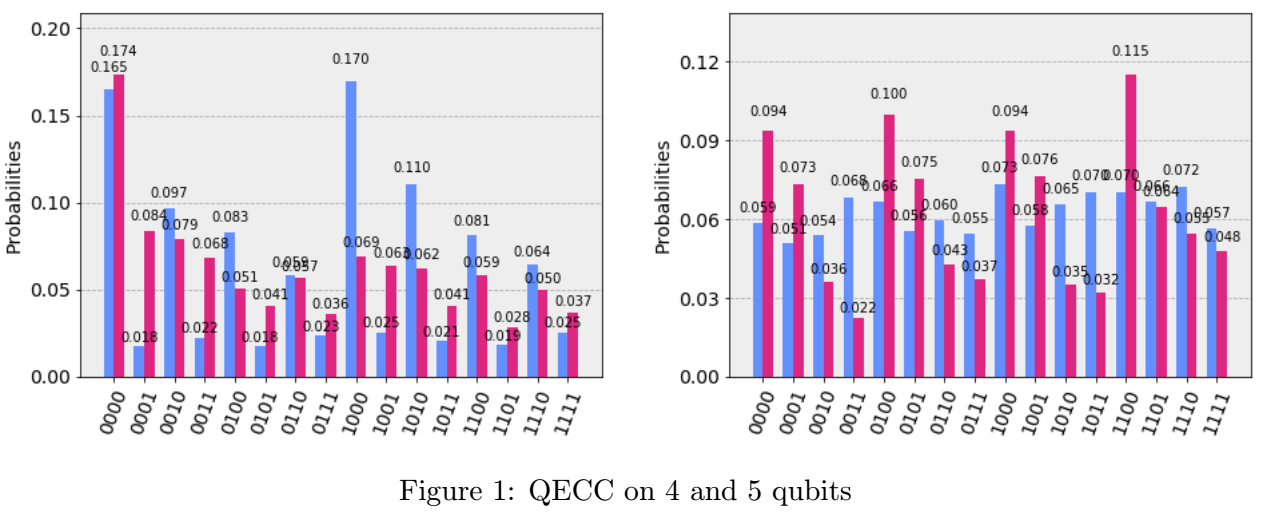}}


\begin{thebibliography}{WWW}

\bibitem{C} M.D. Choi,  Completely positive linear maps on 
complex matrices,
Linear Algebra and Appl. 10 (1975), 285-290.

\bibitem{DS} I. Devetak and P. W. Shor,  The capacity of a 
quantum channel
for simultaneous transmission of classical and quantum 
information, 
Communications in Mathematical Physics,  256, no. 2, pp. 
287–303, 2005.

\bibitem{GLZ} M. Grassl, S. Lu, and B. Zeng, 
Codes for simultaneous transmission of quantum and classical 
information, 
2017 IEEE International Symposium on Information Theory (ISIT),  
1718--1722, 2017.

\bibitem{HW1}  M.H. Hsieh  and M.M. Wilde,  Entanglement-
assisted communication of classical and quantum information, 
IEEE Transactions on Information Theory, 56, no. 9, 4682--4704, 
2010.

\bibitem{HW2} M.H. Hsieh  and M.M. Wilde,  Trading classical 
communication, quantum communication, and entanglement in 
quantum Shannon theory, IEEE Transactions on Information
Theory, vol. 56, no. 9,   4705--4730,   2010.

\bibitem{KL}  
E. Knill, R. Laflamme, and L. Viola, 
Theory of Quantum Error Correction for General Noise,
Physical Review
Letters 84, 2525, 2000.

\bibitem{KBN} Y. Kondo, C. Bagnasco, and M. Nakahara,
Implementation of a simple operator-quantum-error-correction scheme,
Phys. Rev. A 88, 022314, 2013.

\bibitem{Ket}
D.W. Kribs, R. Laflamme, D. Poulin, M. Lesosky, 
Operator quantum error correction,
Quant. Inf. \& Comp., 6, 383-399 (2006)

\bibitem{LNPST2012} 
C.K. Li, M. Nakahara, Y.T. Poon, N.S. Sze and H. Tomita,
Recovery in quantum error correction for general noise without measurement,
Quantum Information \& Computation 12 (2012), 149-158. 
 

\bibitem{LNPST} C.K. Li, M. Nakahara, Y.T. Poon, N.S. Sze and H. Tomita,
Efficient Quantum Error Correction for Fully Correlated Noise, 
Phys. Lett. A, 375:3255-3258, 2011.

\bibitem{MD1} 
M. M\"{u}ller, A. Rivas, E.A. Mart\'{i}nez, D. Nigg, P. Schindler,
T. Mnz, R. Blatt, and M.A. Martin-Delgado,
Iterative Phase Optimization of Elementary Quantum Error Correcting Codes,
Phys. Rev. X 6, 031030, 2016.


\bibitem{Naka} M. Nakahara and T. Ohmi,
Quantum Computing: From Linear Algebra to 
Physical Realizations, CRC Press, New York, 2008.

\bibitem{NC} M.A. Nielsen and I.L. Chuang,
Quantum Computation and Quantum Information, 
Cambridge University Press, Cambridge, 2000.

\bibitem{MD2} D. Nigg, M. M\"{u}ller, E. A. Martinez, P. Schindler, 
M. Hennrich, T. Monz, M. A. Martin-Delgado, R. Blatt,
Quantum computations on a topologically encoded qubit,
SCIENCE Vol. 345 no. 6194 pp. 302-305, 2014.
DOI: 10.1126/science.1253742


\bibitem{Y} J. Yard,  Simultaneous classical-quantum 
capacities of quantum multiple
access channels,  Ph.D. dissertation, Electr. Eng. Dept., 
Stanford Univ., Stanford, CA, 2005.
 
\bibitem{qiskit}
G.  Aleksandrowicz,  T.  Alexander,  P.  Barkoutsos,  L.  
Bello,  Y.  BenHaim,  D.  Bucher,  F.  J.  Cabrera-Hernadez,  
J.  Carballo-Franquis, A.  Chen,  C.-F.  Chen,  J.  M.  Chow,  
A.  D.  Corcoles-Gonzales,  A.  J. Cross,  A.  Cross,  J.  
Cruz-Benito,  C.  Culver,  S.  D.  L.  P.  Gonzalez, E.   D.   
L.   Torre,   D.   Ding,   E.   Dumitrescu,   I.   Duran,   P.   
Eendebak,  M.  Everitt,  I.  F.  Sertage,  A.  Frisch,  A.  
Fuhrer,  J.  Gambetta, B.  G.  Gago,  J.  Gomez-Mosquera,  D.  
Greenberg,  I.  Hamamura, V.  Havlicek,  J.  Hellmers,  Ł.  
Herok,  H.  Horii,  S.  Hu,  T.  Imamichi, T. Itoko, A. Javadi-
Abhari, N. Kanazawa, A. Karazeev, K. Krsulich, P. Liu, Y. Luh, 
Y. Maeng, M. Marques, F. J. Martin-Fernandez, D. T. McClure,  
D.  McKay,  S.  Meesala,  A.  Mezzacapo,  N.  Moll,  D.  M. 
Rodriguez,  G.  Nannicini,  P.  Nation,  P.  Ollitrault,  L.  
J.  O’Riordan, H.  Paik,  J.  Perez,  A.  Phan,  M.  Pistoia,  
V.  Prutyanov,  M.  Reuter, J. Rice, A. R. Davila, R. H. P. 
Rudy, M. Ryu, N. Sathaye, C. Schnabel, E. Schoute, K. Setia, Y. 
Shi, A. Silva, Y. Siraichi, S. Sivarajah, J. A. Smolin, M. 
Soeken, H. Takahashi, I. Tavernelli, C. Taylor, P. Taylour, K.  
Trabing,  M.  Treinish,  W.  Turner,  D.  Vogt-Lee,  C.  
Vuillot,  J.  A. Wildstrom, J. Wilson, E. Winston, C. Wood, S. 
Wood, S. Worner, I. Y. Akhalwaya, and C. Zoufal, “Qiskit: An 
open-source framework for quantum computing,” 2019.

\bibitem{ERR}
 \verb|https://www.research.ibm.com/ibm-q/technology/devices/|
 
\end{thebibliography}
\end{document}